\documentclass{llncs}

\usepackage{lipsum} 

\usepackage[utf8]{inputenc}
\usepackage[T1]{fontenc} 
\usepackage{microtype} 
\usepackage{tablefootnote}
\usepackage{mathtools}
\usepackage{amssymb}
\usepackage{microtype}
\usepackage{algorithm}
\usepackage{algpseudocode}
\usepackage{marginnote}
\usepackage[normalem]{ulem}

\usepackage{color}
\usepackage{cleveref}
\usepackage{paralist} 
\usepackage{booktabs}

\newcommand{\bigoh}{\mathcal{O}}

\usepackage{thmtools, thm-restate}

\begin{document}

\title{A linear time algorithm to compute\\ the \emph{impact} of all the articulation points}
\author{Gabriele Farina$^1$}
\footnotetext[1]{Polytechnic University of Milan, Italy. E-mail: \email{gabr.farina@gmail.com}}
\institute{}
\maketitle
\thispagestyle{empty}

\vspace{-.3cm}

\begin{abstract}
\noindent The articulation points of an undirected connected graph are those vertices whose removal increases the number of connected components of the graph, i.e. the vertices whose removal disconnects the graph. However, not all the articulation points are equal: the removal of some of them might end in a single vertex disconnected from the graph, whilst in other cases the graph can be split in several small pieces. In order to measure the effect of the removal of an articulation point, Ausiello et al. \cite{AFL12} have proposed the \emph{impact}. Impact is defined as the number of vertices that get disconnected from the main (largest) surviving connected component (CC) after the removal of the articulation point. 
In this paper we present the first linear time algorithm ($\bigoh(m+n)$ for a graph with $n$ vertices and $m$ edges) to compute the impact of all the articulation points of the graph, thus improving from the $\bigoh(a(m+n))\approx\bigoh(nm+n^2)$ of the na\"ive algorithm, with $a$ being the number of articulation points of the graph. 

\end{abstract}

\vspace{-.8cm}


\section{Introduction}

Connectivity is one of the most basic structural properties of a graph. If an undirected graph is connected, a natural related question is which are the vertices whose removal disconnects the graph; this question has been answered by Hopcroft and Tarjan in \cite{HT73}, that proposed the now classical linear time algorithm, described in many textbooks (see, e.g., \cite{CLRS09}), based on a Depth First Search visit of a graph \cite{Tar72}.

From the topological point of view all the articulation points are equal, in the sense that they split the graph in more than one connected components; however, it is natural, especially when studying the resiliency of a graph, to ask which are the vertices (i.e., articulation points) whose removal is more disrupting. In order to measure the effect of the removal of an articulation point, Ausiello et al. \cite{AFL12} have proposed the \emph{impact}. Impact is defined as the number of vertices that get disconnected from the main (largest) surviving connected component (CC) after the removal of the articulation point. Consider, as an example, the graph shown in Figure~\ref{fig:graphccs}: in this graph the vertex $4$ is the one with the largest impact, and its removal disconnects six vertices from largest connected component (i.e, the one that includes vertices $\{5,6,7,8,9,10\}$).

From its definition, the \emph{impact} can be na\"ively computed in the following way: we first compute, using Hopcroft and Tarjan's algorithm, the articulation point of the graph, then, we remove each of them, one at a time, and perform any linear visit of the graph, i.e. a DFS or a BFS, to compute the number of reachable vertices. If the graph has $a$ articulation points, this na\"ive algorithm costs $\bigoh(a(m+n))\approx\bigoh(nm+n^2)$, since $a$ can be of the same order of $n$, and indeed it holds $0\leq a \leq n-2$.

In this paper we present the first linear time algorithm to compute the impact for \emph{all} the articulation points of an undirected graph in linear time, i.e. $\bigoh(m+n)$ for a graph with $n$ vertices and $m$ edges. The algorithm is built over three main ingredients: the block forest data structure, proposed by Westbrook and Tarjan in \cite{WT92} to maintain the biconnected components of a graph in an online setting, an algorithm to recursively compute this data structure offline, and a depth first based traversal of this structure, to compute the \emph{impact} of all the vertices. 

This paper is organized as follows: the few necessary preliminary notions are discussed in Section~\ref{sec:prelim}. The algorithm is described in Section~\ref{sec:alg}, whilst concluding remarks are addressed in Section~\ref{sec:conclusions}. Due to space constraints, we provide only some intuition of the algorithm properties, and do not prove them formally. In the Appendix we present the algorithm's pseudocode and some other remarks.



\section{Preliminaries.}
\label{sec:prelim}
Given an undirected graph $G=(V,E)$, we define:
\begin{description}
	\item[Connected component:] a maximal set of vertices $V'\subseteq V$ such that, given $u,v\in V'$, there is at least one path between $u$ and $v$ in $G$. 
	\item[Articulation point:] a vertex $v\in V$ such that its removal from the graph $G$ increases the number of connected components in $G$. 
	\item[Bridge:] an edge $e\in E$ such that its removal from the graph $G$ increases the number of connected components in $G$. 
	\item[Biconnected component:] a maximal set of vertices $V''\subseteq V$ such that the removal of any single vertex $v\in V''$ leaves the component connected.
\end{description}

\begin{figure}[t!]
	\includegraphics[scale=.51]{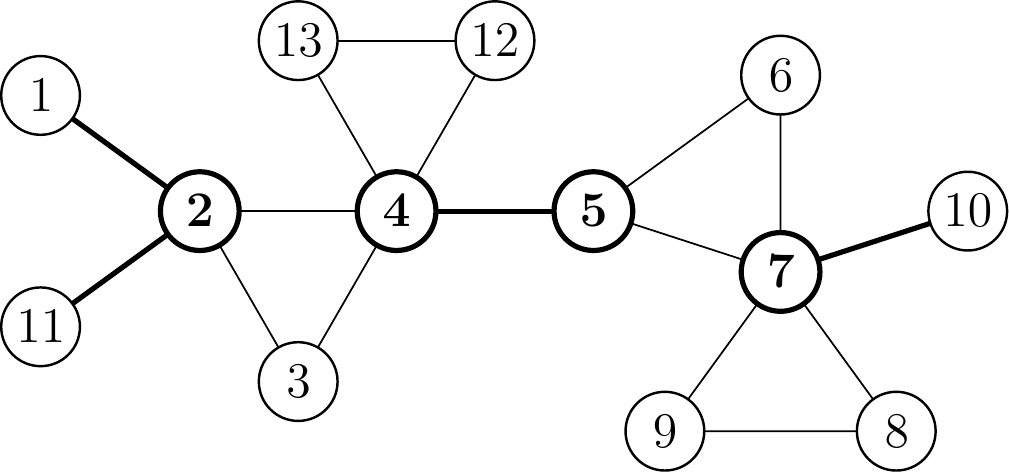}\hfill \raisebox{1cm}{$\longrightarrow$} \hfill
	\includegraphics[scale=.51]{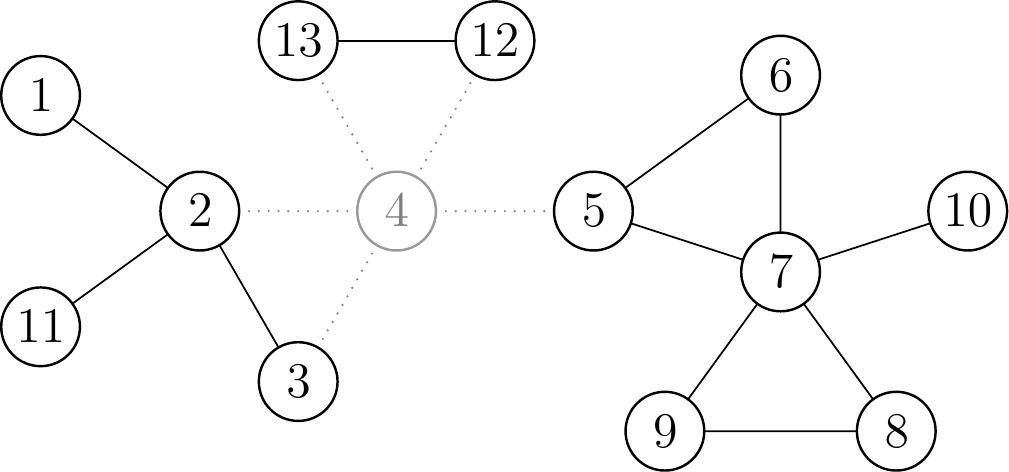}
	\caption{Example graph. Articulation points (bold vertices), and bridges (bold edges) are emphasized [left]; the effect of the removal of vertex $4$ [right].\label{fig:graphccs}}
\end{figure}

In Figure~\ref{fig:graphccs} [left] we can see an example graph arranged in order to emphasize its articulation points and bridges. In Figure~\ref{fig:bfds} [left] we can see the same graph, with its biconnected components emphasized. It is important to note that all the vertices adjacent to bridges are articulation points unless the bridge is their only incident edge, as is the case for vertex $1$ in Figure~\ref{fig:graphccs}: the removal of the bridge leaves vertex $1$ isolated, but the removal of vertex $1$ (together with all its adjacent edges, i.e. only the bridge) does not increase the number of connected components of $G$.

\section{The algorithm}
\label{sec:alg}

\begin{figure}[t!]
	\centering \raisebox{0.6cm}{ \includegraphics[scale=.49]{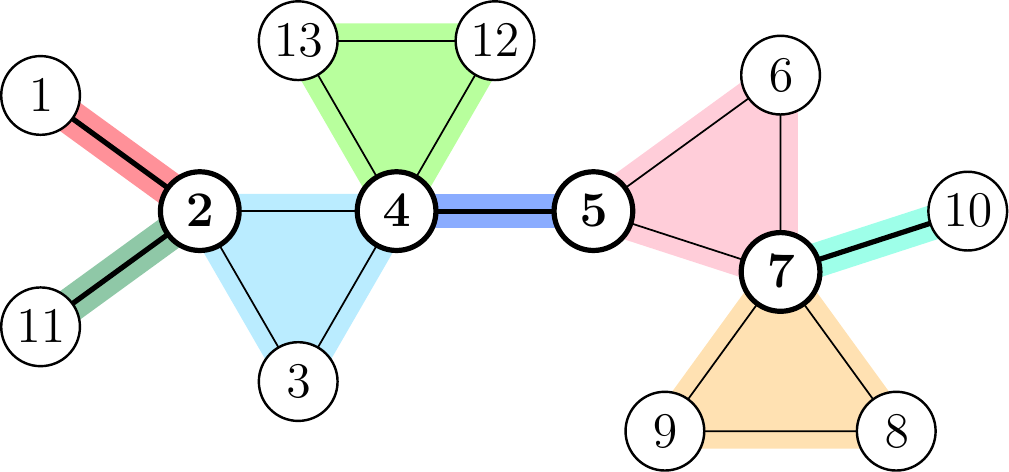}}\hfill 	\includegraphics[scale=.50]{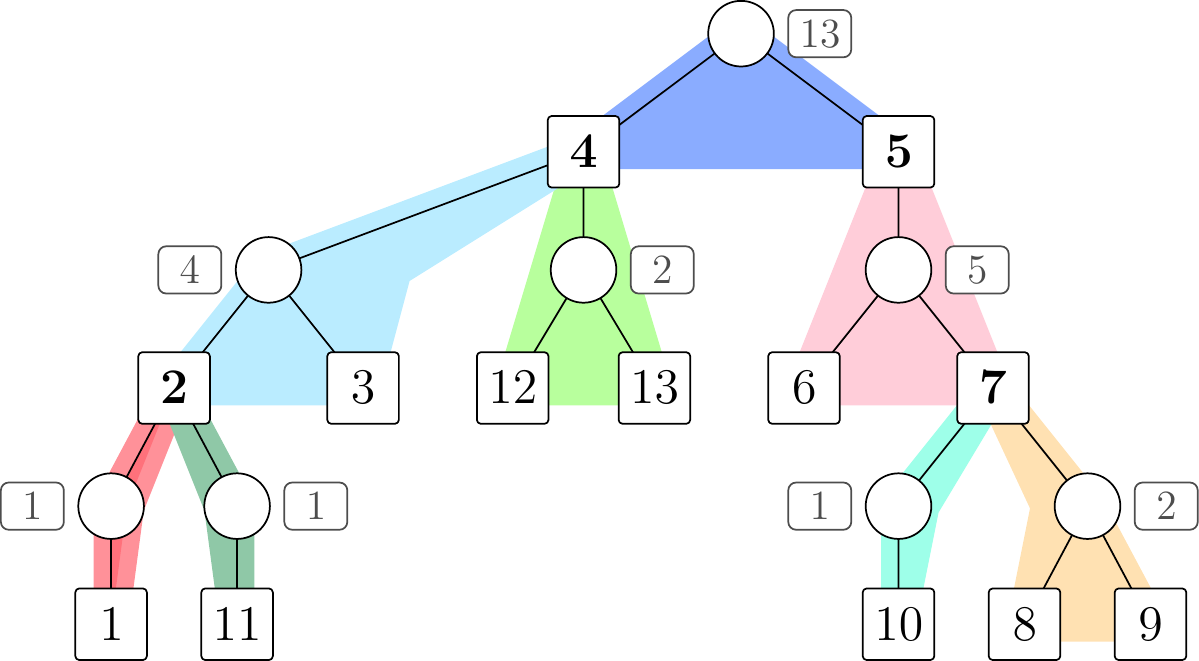}
	\caption{Biconnected components of the graph of Figure~\ref{fig:graphccs} [left] and the corresponding block forest data structure [right].\label{fig:bfds}}
\end{figure}

In this section we describe our linear time algorithm to compute the impact of all the vertices (articulation points). As we mentioned in the introduction, the algorithm uses the block forest (BF) data structure, proposed by Westbrook and Tarjan in \cite{WT92} to maintain online the biconnected components. In order to provide some intuition for the algorithm we propose, we briefly recall some properties of the block forest data structure. In Figure~\ref{fig:bfds} it is possible to see the graph from Figure~\ref{fig:graphccs} and the corresponding block forest data structure that, since the graph is connected, is a single tree. This tree has two distinct type of vertices: square vertices, that correspond to vertices of the original graph, and round vertices, that correspond to the biconnected components of the graph. Each  vertex of the graph (square vertices) is connected, in the BF, to all the biconnected components it belongs to (round vertices). Every tree in the block forest is rooted in a round node, unless the tree represents a singleton (i.e., a vertex with no incident edges) of the graph, in which case the tree is made of a single square node. The articulation points of the graph, shown in figure as bold square vertices, are the square vertices that are not leaves in the BF. 

Now that we described the properties of the BF data structure, the algorithm can be roughly divided into two main steps:
\begin{itemize}
\item it computes the BF data structure offline, using a recursive offline algorithm;
\item it computes the impact of each (square non-leaf) vertex in the BF, using a DFS based algorithm (see the pseudocode in the appendix).
\end{itemize}

We provide some intuition for both the steps of the algorithm: the first one builds the BF data structure one tree at a time, starting from an unvisited vertex of the graph at each step; the construction of a single tree can be seen as a modified version of Hopcraft and Tarjan's algorithm \cite{HT73}. In order to compute the impact of the vertices, every round vertex of the BF stores the number of square vertices that belong to its subtree. This can be seen graphically in the BF tree shown in Figure~\ref{fig:bfds}, where these numbers are put in the small gray badges close to the round vertices. It can be proven that the impact, for each square vertex, can be easily calculated by looking at the values written in its round neighbours. We compute these values for each square vertex using a DFS based algorithm.

\section{Conclusions}
\label{sec:conclusions}

In this paper we presented a linear time algorithm to compute, in linear time, the \emph{impact} of all the articulation points of an undirected graph. More precisely, we showed that the following holds:
\begin{theorem}
	Given an undirected graph $G = (V, E)$, it is possible to compute the impact of all the articulation points in $\bigoh(|V| + |E|)$ time and space.
\end{theorem}
The approach described in this paper could be extended to directed graphs as well but, even though recently a linear time algorithm has been proposed to compute all the strong articulation points, it is still open the problem of how to compute the 2-vertex connected components \cite{ILS12}.


\bibliographystyle{abbrv}
\bibliography{impact}
\vfill 
\end{document}